
%
\input harvmac
\def\ack{\bigbreak\bigskip\bigskip\centerline{{\bf Acknowledgements}}\nobreak}
\baselineskip=16pt plus 2pt minus 1pt
\newskip\normalparskip
\normalparskip = 6pt plus 2pt minus 1pt
\parskip = \normalparskip
\parindent=12pt

\def\a{\alpha}    \def\b{\beta}       \def\c{\chi}       
    \def\e{\varepsilon}        
    \def\G{\Gamma}           \def\l{\lambda}
   \def\m{\mu}         \def\n{\nu}        
\def\vr{\varrho}  \def\o{\omega}      \def\O{\Omega}     \def\p{\psi}
      \def\s{\sigma}      \def\S{\Sigma}

\def\CG{{\cal G}}
\def\CM{{\cal M}}

\def\CD{{\cal D}}

%
\font\teneufm=eufm10
\font\seveneufm=eufm7
\font\fiveeufm=eufm5
\newfam\eufmfam
\textfont\eufmfam=\teneufm
\scriptfont\eufmfam=\seveneufm
\scriptscriptfont\eufmfam=\fiveeufm
\def\eufm#1{{\fam\eufmfam\relax#1}}

\font\teneusm=eusm10
\font\seveneusm=eusm7
\font\fiveeusm=eusm5
\newfam\eusmfam
\textfont\eusmfam=\teneusm
\scriptfont\eusmfam=\seveneusm
\scriptscriptfont\eusmfam=\fiveeusm

\font\tenmsx=msam10
\font\sevenmsx=msam7
\font\fivemsx=msam5
\font\tenmsy=msbm10
\font\sevenmsy=msbm7
\font\fivemsy=msbm5
\newfam\msafam
\newfam\msbfam
\textfont\msafam=\tenmsx  \scriptfont\msafam=\sevenmsx
  \scriptscriptfont\msafam=\fivemsx
\textfont\msbfam=\tenmsy  \scriptfont\msbfam=\sevenmsy
  \scriptscriptfont\msbfam=\fivemsy

\def\msbm#1{{\fam\msbfam\relax#1}}



\def\darr#1{\raise1.5ex\hbox{$\leftrightarrow$}\mkern-16.5mu #1}
\def\Ha{{1\over2}}

\def\Fr#1#2{{#1\over#2}}
\def\tr{\hbox{Tr}\,}

\def\Fs#1{#1\!\!\!\!/\,} 
\def\roughly#1{\raise.3ex\hbox{$#1$\kern-.75em\lower1ex\hbox{$\sim$}}}

%
\def\cmp#1#2#3{Comm.\ Math.\ Phys.\ {{\bf #1}} {(#2)} {#3}}
\def\pl#1#2#3{Phys.\ Lett.\ {{\bf #1}} {(#2)} {#3}}
\def\np#1#2#3{Nucl.\ Phys.\ {{\bf #1}} {(#2)} {#3}}

\def\jdg#1#2#3{J.\ Differ.\ Geom.\ {{\bf #1}} {(#2)} {#3}}

\def\top#1#2#3{Topology {{\bf #1}} {(#2)} {#3}}


\def\pr{\prime}

\def\adP{\eufm{g}_{\raise-.1ex\hbox{${}_P$}}}
\def\gBE{\eufm{g}_{\raise-.1ex\hbox{${}_\BE$}}}
\def\gEc{\eufm{g}_{\raise-.1ex\hbox{${}_E$}}^C}
\def\BC{\msbm{C}}
\def\BE{\msbm{E}}

\def\BR{\msbm{R}}
\def\BZ{\msbm{Z}}

\lref\Donaldson{
S.~Donaldson,
\top{29}{1990}{257}.
}
\lref\DonaldsonB{
S.~Donaldson,
\jdg{26}{1987}{397}.
}
\lref\WittenA{
E.~Witten,
\cmp{117}{1988}{353}.
}
\lref\WittenB{
E.~Witten,
J.~Math.~Phys.~{\bf 35} (1994) 5101.
}
\lref\WittenC{
E.~Witten,
Math.~Research Lett.~{\bf 1} (1994) 769.
}
\lref\WittenD{
E.~Witten,
\np {B 371}{1992}{191}; Mirror manifolds and topological field theory,
in {\it Essays on mirror manifolds,}
ed.~S.-T.~Yau (International Press, 1992).
}
\lref\WittenE{
E.~Witten,
\np{B 403}{1993}{159}.
}
\lref\WittenN{
E.~Witten, Lectures given at the Isaac Newton Institute, Dec.~1994.
}
\lref\VW{
C.~Vafa and E.~Witten,
\np{B 431}{1994}{3}.
}
\lref\SWa{
N.~Seiberg and E.~Witten,
\np{B 426}{1994}{19}.
}
\lref\SWb{
N.~Seiberg and E.~Witten,
\np{B 431}{1994}{484}.
}
\lref\MO{
C.~Montonen and D.~Olive,
\pl{B 72}{1977}{117}.
}
\lref\WO{
E.~Witten and D.~Olive,
\pl{B 78}{1978}{97}.
}
\lref\KM{
P.~Kronheimer and T.~Mrowka,
The genus of embedded surfaces in the projective plane,
preprint, 1994.
}
\lref\MR{
T.~Mrowka, private communication.
}
\lref\Taubes{
C.~Taubes,
Symplectic manifolds and the Seiberg-Witten invariants,
preprint, 1994.
}
\lref\HPPa{
S.~Hyun, J.~Park and J.-S.~Park,
Topological QCD on a K\"{a}hler manifold,
to appear.
}
\lref\HPPb{
S.~Hyun, J.~Park and J.-S.~Park,
N=2 supersymmetric QCD and four manifolds (I), (II).,
to appear.
}
\lref\WB{
J.~Wess and J.~Bagger, Supersymmetry and
Supergravity, second edition (Princeton University Press, 1992).
}
\lref\PT{
V.~Pidstrigach and A.~Tyurin,
Izv.~AN SSSR Ser.~Math.~56:2 (1992) 279
(English translation: AMS {\bf 40} (1993) 163).
}
\lref\TyurinA{
A.~Tyurin,
The simple method of distinguishing the underlying differential
 structures of
algebraic surfaces,
alg-geom/9210003;
Spin canonical invariants of $4$-manifolds and algebraic surfaces,
alg-geom/9406002.
}
\lref\PG{V.~Pidstrigach, Lecture given at Isaac Newton Inst. (Dec.~1994).
}
\lref\TyurinB{A.~Tyurin, Lecture given at Isaac Newton Inst. (Dec.~1994).
}
\lref\GSW{
M.B.~Green, J.H.~Schwartz and E.~Witten,
Superstring theory, Vol.~2,
(Cambridge University Press, 1987).
}
\lref\ML{
H.B.~Lawson, JR and M-L. Michelsohn,
Spin geometry, (Princeton University Press,1989).
}
\lref\Yamaron{
J.~Yamaron,
\pl{B 213}{1988}{325}.
}
\lref\AF{
P.~Argyres and A.~Faraggi,
The Vacuum structure and spectrum of $N=2$ supersymmetric
$SU(N)$ gauge theory,
hep-th/9411057.
}
\lref\KLYT{
A.~Klemm, W.~Lerche, S.~Yankielowicz and S.~Theisen,
Simple singularities and $N=2$ supersymmetric Yang-Mills theory,
hep-th/9411048;
On the monodromies of $N=2$ supersymmetric Yang-Mills theory,
hep-th/9412158.
}
\lref\CMR{
S.~Cordes, G.~Moore and S.~Ramgoolam,
Lectures on 2D Yang-Mills theory, equivariant cohomology and
topological field  theories, Part II, hep-th/9411210.
}

\lref\ANFa{
D.~Anselmi and P.~Fr\'{e},
Gauged hyperinstantons and monopole equations,
hep-th/9411205.
}
\lref\ANFb{
D.~Anselmi and P.~Fr\'{e},
\np{B 404}{1993}{288};
\np{B 416}{1994}{255}.
}
\lref\LM{
J.M.F.~Labastida and M.~Mari$\tilde{\rm n}$o,
A topological Lagrangian for monopoles on four-manifolds,
hep-th/9503105.
}
\lref\FAYA{
P.~Fayet,
\np{B 113}{1976}{135}
}
\lref\SHW{
P.S.~Howe, K.S.~Stelle and P.C.~West,
\pl{B 124}{1983}{55}.
}
\lref\SBa{
N.~Seiberg,
\pl{B 206}{1988}{75}.
}
\lref\SBb{
N.~Seiberg,
Electro-magnetic duality in supersymmetric non-abelian gauge theories,
hep-th/9411149.
}
\lref\SBc{
N.~Seiberg,
The power of holomorphy - exact results in $4$d SUSY field theories,
hep-th/9408013 and references therein.
}
\lref\HP{
S.~Hyun and J.-S.~Park,
Holomorphic Yang-Mills theory and the variation of
the Donaldson invariants,
hep-th/9503036.
}
\font\Titlerm=cmr12 scaled\magstep3
\nopagenumbers
\rightline{YUMS-95-08, CALT-68-1985,  SWAT/67}
\rightline{hep-th/9503201}
\vskip .5in
\centerline{\fam0\Titlerm  TOPOLOGICAL QCD}
\tenpoint\vskip .4in\pageno=0
\centerline{
Seungjoon Hyun
}
\medskip
\centerline{{\it Institute for Mathematical Sciences, Yonsei University}}
\centerline{{\it Seoul 120-749, Korea}}
\centerline{{\it (hyun@phya.yonsei.ac.kr)}}
\medskip
\centerline{and}
\medskip
\centerline{
Jaemo Park
}
\medskip
\centerline{\it Department of Physics, California Institute of Technology}
\centerline{\it Pasadena, CA 91125, USA}
\centerline{\it (jaemo@cco.caltech.edu)}
\medskip
\centerline{and}
\medskip
\centerline{
Jae-Suk Park\footnote{$^{\dagger}$}{
Address after 1 Oct.~1995 : Institute for Theoretical Physics, University of
Amsterdam,
Valekenierstraat 65, 1018 XE Amsterdam.
}
}
\medskip
\centerline{\it Department of Physics, University of Wales, Swansea}
\centerline{\it Swansea SA2 8PP, UK }
\centerline{\it (j.park@swansea.ac.uk)}
\bigskip

\noindent\abstractfont
We study  the twisted  $N=2$ supersymmetric
Yang-Mills theory  coupled with the hypermultiplets (TQCD).
We suggest that the family of TQCD  can be served as a powerful  tool
for studying the quantum field theoretic  properties of the underlying
physical  theories.

\Date{March, 1995;  to appear in Nucl.~Phys.~B}


\newsec{Introduction}
The minimal $N=2$ super-Yang-Mills (SYM in short) theory
can be twisted to define the topological Yang-Mills (TYM in short)
 theory \WittenA.
The underlying asymptotically free physical theory has two
different limits;
the weakly coupled ultraviolet and the strongly coupled infrared
limits.
The path integrals of the twisted theory
can be computed in both limits corresponding to the
two different scaling limits.
The ultraviolet limit of the theory gives the original cohomological
description  of the Donaldson invariants \Donaldson.
On the other hand, the infrared limit
of the theory gives an entirely different viewpoint of the Donaldson
invariants.
This physical realization of the Donaldson invariants provides a
powerful and genuine new viewpoint on the invariants of the
smooth structure on  four-manifolds \WittenB\WittenC.

Recently,  Seiberg and Witten studied the strongly coupled infrared
limit of
the minimal $N=2$ SYM theory \SWa. Based on the resulting
low-energy effective theory, Witten introduced new four-manifold
invariants (the Seiberg-Witten invariants), which is the dual
description of the Donaldson invariants, and determined the
Donaldson invariants completely for  K\"{a}hler
surfaces with $b_2^+ \geq 3$ \WittenC.\foot{The variation of the
Donaldson
invariants on K\"{a}hler surface with
$b_2^+ =1$ is studied in \HP.}
This new approach turns out to be extremely powerful in many
 respects \KM\Taubes.
The Seiberg-Witten monopole theory can be viewed as
the twisted version\foot{This is not rigorously true as we shall
see later.}
of the $N=2$ super-Maxwell theory with
one massless hypermultiplet \WittenC, which arises as  the
low energy
effective theory of the minimal $N=2$ SYM
theory \SWb.  It is very surprising that the abelian theory
with matter describes the highly non-trivial four-manifold
invariants after twisting.

In their second paper \SWb, Seiberg and Witten studied the
strongly coupled infrared limit of the  $N=2$ supersymmetric
$SU(2)$ Yang-Mills theories coupled with hypermultiplets
in the fundamental representation.
An interesting new feature, besides from the many physical
 implications,
is that there exists a critical theory with exactly vanishing
$\beta$-function.
The asymptotically free non-critical theories seem to belong to
 the same universality class  as certain massive deformations
 of the critical theory.
Furthermore, the critical theory has shown to have the
full $SL(2,\BZ)$ symmetry exchanging electric and magnetic
charges analogous to Montonen-Olive duality \MO\WO, as refined
and tested
in \VW, of the $N=4$ theory. One can naturally believe the above
features are general properties of the $N=2$ supersymmetric
gauge theories.
This leads to the program of the classification of the theory up
to the universality class and the duality.

In this paper, we study  the
twisted $N=2$ SYM theories coupled with the hypermultiplets on
a compact oriented  simply connected\foot{
The restriction for the simply connected case is just for
convenience.
A detailed account for  K\"{a}hler surface will be
discussed in our forthcoming paper \HPPa.}
Riemann $4$-manifold $X$. The crucial observation is
that the twisting procedure in general (including the
minimal case) should always involve the $spin^c$ structure on $X$.
The appearance of
the $spin^c$ structure in the topological twisting of the $N=2$ SYM
theory coupled with the $N=2$ matters is a special new feature which
adds very rich flavors to the TQCD.
The other crucial property is, as expected,
that the theory without matter is independent of the choice
of the $spin^c$ structure used in the twisting.
We refer to the theory as the topological QCD (TQCD) and
the resulting topological invariants as the
Donaldson-Seiberg-Witten (DSW)
invariants or the monopole invariants.
This invariants have some similarities with the
spin polynomials  proposed by Pidstrigach and Tyurin \PT\TyurinA\
who first used non-abelian $spin^c$ Dirac operator to define
smooth invariants.

For a given compact connected semi-simple gauge group,
we have  a  family of the monopole invariants
associated with arbitrary
representations, whose data define the underlying
asymptotically free or scale
invariant theory,
as well as with arbitrary $spin^c$ structures on $X$.
Of course the cohomological definition of the DSW invariants
is based on the weakly coupled ultraviolet limit of the underlying $N=2$
supersymmetric QCD.
More importantly,
we suggest that the family of TQCD  can be served as  powerful
new tool for studying the quantum field theoretic properties of
the
underlying physical  theories.

This paper is organized as follows;
In sect.~2,  we review  the  $N=2$ supersymmetric Yang-Mills
theory coupled
with hypermultiplets.
In Sect.~3,  we twist the hypermultiplets and define
the topological  QCD.
We discuss the meaning of fermionic zero-modes
and calculate the ghost number anomaly (the dimension of
the moduli space).
In Sect.~4, we study the topological observables and
their correlation functions which define the DSW invariants.
Finally we add some remarks on the the original Seiberg-Witten monopole
equation and on the future prospects of the TQCD in Sect.~5.
Up to Sect.~4, we restrict our attentions to the theory
with one hypermultiplet.
In the final section, we will briefly consider more general cases and their
possible applications to the underlying physical theories.

Many properties of TQCD as a cohomological theory can be
trivially established by adopting the similar arguments
of Witten on  TYM theory. Hence, we will freely refer to
the original paper of Witten \WittenA. This paper has been grown out from
our efforts to understand the Seiberg-Witten theory \WittenC\SWa.
We are also motivated by Witten's suggestion
that one can generalize the Seiberg-Witten equation to
define a new family of four-manifold invariants \WittenN.

We should note that the twisting of $N=2$ hypermultiplets with arbitrary
gauge group was previously studied by  Anselmi and Fr\'{e} \ANFb.\foot{
We are grateful to D.~Anselmi for pointing this out after
the first version of this paper.}
They use the $\s$-model interpretation of spin-$0$ bosons of a
hypermultiplet while we use $spin^c$ structures in twisting those
bosons. However, the resulting equations obtained from the two approaches
coincide with each other for a hyperk\"{a}hler manifold \ANFa.
Furthermore, they worked out the twisting of the most general $N=2$
supersymmetric theories including  gravity \ANFb.
After finishing this paper, a paper on
a construction of topological action  for the abelian Seiberg-Witten
monopoles appeared \LM.

\newsec{The $N=2$ Supersymmetric QCD}
\subsec{The Physical Action and Supersymmetry}

To begin with, we briefly review the $N=2$ supersymmetric Yang-Mills
theory coupled with hypermultiplets.  We generally follow the notation
of  Wess and Bagger \WB\
as well as Seiberg and Witten's papers \SWa\SWb.
We consider a compact connected simple gauge group $G$.
We adopt the anti-hermitian convention of the Lie algebra generators.

The Lorentz symmetry $K$ is locally
isomorphic to $SU(2)_L\times SU(2)_R$. In addition to $K$,
the $N=2$ SYM theory may have the internal symmetry
$U(2)_I$. The instanton breaks $U(2)_I$ to a subgroup whose connected
component  is isomorphic  to $SU(2)_I\times U(1)_\CR$.
$U(1)_\CR$ is anomalous unless the theory is scaling (conformal) invariant.
Its charge is denoted as a quantum number $U$.
The  global symmetry of the theory is
\eqn\adb{
H = SU(2)_L\times SU(2)_R\times SU(2)_I.
}
The supercharges $Q_{\a}{}^i$ and $Q_{\dot{\a}i}$ transform under $H$ as
$(\Ha,0,\Ha)^{-1}$ and  $(0,\Ha,\Ha)^{+1}$,
where the superscript denotes the $U$ charge.
The minimal action consists of the $N=2$ vector multiplets
in the adjoint representations;
\eqn\adc{\matrix{
 &\quad         &A_m\quad &{}\cr
 &\l\quad       &\phantom{\psi^q}\quad &\psi\cr
 &\quad         &B \quad &{}\cr
}
}
Note that $\l =\l_{\a}^1$, $\psi=\l_{\a}^2$,
$\bar{\l}=\bar{\l}^{\dot \a}_1$
and $\bar{\psi}=\bar{\l}^{\dot \a}_2$.
The quantum numbers for the fields under $H$ are
\eqn\aad{\eqalign{
\l &= (\Ha,0,\Ha)^{+1},\cr
\bar{\l}&=(0,\Ha,\Ha)^{-1},\cr
}\qquad
\eqalign{
\psi &= (\Ha,0,\Ha)^{+1},\cr
\bar{\psi} &= (0,\Ha,\Ha)^{-1},\cr
}\qquad
\eqalign{
B &= (0,0)^{+2},\cr
\bar B &=(0,0)^{-2}.\cr
}
}

The $N=2$ supersymmetric action is given by
\eqn\aaa{\eqalign{
h^2 L_{YM}=& -\Fr{1}{4}F_{mn}^{a}F^{mn}_{a}
-i\bar{\lambda}_{\dot{\alpha} i}^{a}
\bar{\sigma}^{m \dot{\alpha}\alpha}D_{m}\lambda_{\alpha a}{}^{i}
-D_{m}\bar{B}^{a}
D^{m}B_{a}
\cr
& -\Fr{i}{\sqrt{2}}\lambda^{\alpha ia}[\bar{B},\lambda_{\alpha i}]_{a}
- \Fr{i}{\sqrt{2}}\bar{\lambda}_{\dot{\alpha}}{}^{ia}
[B,\bar{\lambda}^{\dot{\alpha}}{}_{i}]_{a}-\Fr{1}{2}[\bar{B},B]^2,\cr
}
}
where $h^2$ denotes the coupling constant.
The supersymmetry transformation is
\eqn\aab{\eqalign{
\delta A_{m}=
&i\xi^{\alpha i}\sigma_{m\alpha\dot{\alpha}}
\bar{\lambda}^{\dot{\alpha}}{}_{i}
-i\lambda^{\alpha
i}\sigma_{m\alpha\dot{\alpha}}\bar{\xi}^{\dot{\alpha}}{}_{i},
     \cr
\delta\lambda_{\alpha}{}^{i}=
&\sigma^{mn \, \beta}_{\,\,\alpha}\xi_{\beta}{}^{i}F_{mn}
    +\sqrt{2}i\sigma^{m}_{\alpha\dot{\alpha}}D_{m}
    B \bar{\xi}^{\dot{\alpha} i}
    +[B,\bar{B}] \xi_{\alpha}{}^{i},
     \cr
\delta\bar{\lambda}_{\dot{\alpha} i} =
&-\bar{\xi}_{\dot{\beta} i}
\bar{\sigma}^{mn\dot{\beta}}_{\quad \,\dot{\alpha}}
F_{mn}
   +\sqrt{2}i\xi^{\alpha i}\sigma^{m}_{\alpha \dot{\alpha}}D_{m}\bar{B}
   -[B,\bar{B}]\bar{\xi}_{\dot{\alpha} i},
     \cr
\delta B=
&\sqrt{2}\xi^{\alpha i}\lambda_{\alpha i},
      \cr
\delta\bar{B}=
&\sqrt{2}\bar{\xi}^{\dot{\alpha}}_{i}\bar{\lambda}_{\dot{\alpha}}{}^{i},
      \cr
}
}

In addition to the vector multiplets one can couple
the $N=2$ supersymmetric
matter fields called the hypermultiplets \FAYA.\foot{In the coupled
theory the transformation law \aab\ for the $N=2$ vector multiplet
should be changed as
$$\eqalign{
\delta\lambda_{\alpha}{}^{i a}=
&\sigma^{mn \, \beta}_{\,\,\alpha}\xi_{\beta}^{i}F_{mn}^{a}
    +\sqrt{2}i\sigma^{m}_{\alpha\dot{\alpha}}(D_{m}B)^{a}
    \bar{\xi}^{\dot{\alpha} i}
    +[B,\bar{B}]^{a} \xi_{\alpha}{}^{i}
    +q^{i\dagger}T^{a}q^{j}\xi_{\alpha j}-q_{j}^{\dagger}T^{a}q^{i}
      \xi_{\alpha}{}^{j},
     \cr
\delta\bar{\lambda}_{\dot{\alpha} i}^{a} =
&-\bar{\xi}_{\dot{\beta} i}
\bar{\sigma}^{mn\dot{\beta}}_{\quad \,\dot{\alpha}}
F_{mn}^{a}
   +\sqrt{2}i\xi^{\alpha i}\sigma^{m}_{\alpha \dot{\alpha}}(D_{m}\bar{B})^{a}
   -[B,\bar{B}]^{a}\bar{\xi}_{\dot{\alpha}i}
   +q_{i}^{\dagger}T^{a}q^{j}\bar{\xi}_{\dot{\alpha}j}
  -q_{j}^{\dagger}T^{a}q_{i}\bar{\xi}_{\dot{\alpha}}{}^{j}.
     \cr
}
$$
We will not use this transformation law in twisting and introduce an auxiliary
field. If one twists the above transformation law one directly get the
twisted transformation law  that can be obtained
after eliminating the auxiliary field,
which will be discussed in the
next section.
}
To couple the hypermultiplet, one
should specify a representation $R$ of
 the gauge group $G$.\foot{
We denotes  $T_{a}$ as the generator of the gauge group in the
representation $R$,
and $[T_{a}, T_{b}]=f^{abc}T_{c}$.}
For simplicity, consider only one hypermultiplet.
A hypermultiplet consists of two Weyl fermions
$\psi_{\!q}$ and $\psi_{\!\tilde{q}}^\dagger$ and complex bosons
$q$ and $\tilde q^\dagger$;
\eqn\ade{\matrix{
 &\quad         &\psi_q\quad &{}\cr
 &q\quad         &\phantom{\psi^q}\quad &{\tilde q}^{\dagger}\cr
 &\quad         &{\psi}_{\tilde{q}}^\dagger\quad &{}\cr
}
}
The quantum numbers of the fields under $H$ are
\eqn\aae{\eqalign{
\psi_{q\a } & = (\Ha,0,0)^{-1}, \cr
\bar{\psi}_q^{\dot \a} &= (0,\Ha,0)^{+1}\,\cr
q&= (0,0,\Ha)^{0},\cr
\tilde q^{\dagger} & =(0,0,\Ha)^{0},\cr
}\qquad\eqalign{
\bar{\psi}_{\tilde q}^{\dot \a} &= (0,\Ha,0)^{+1} ,\cr
\psi_{{\tilde q} \a} &= (\Ha,0,0)^{-1} ,\cr
q^{\dagger}&= (0,0,\Ha)^{0},\cr
\tilde q & =(0,0,\Ha)^{0},\cr
}
}
where
\eqn\aaf{
q^{1} \equiv q,\qquad
q^{2} \equiv \tilde{q}^{\dagger},\qquad
q_{1}^{\dagger}=q^{\dagger},\qquad
q_{2}^{\dagger}=\tilde{q},
}
and $q_{1}=\epsilon_{12}q^{2}=-q^{2}$,
$q_{2}=\epsilon_{21}q^{1}=q^{1}$.

The Lagrangian for the matter fields is given by
\eqn\aag{
\eqalign{
h^2 L_{Matter} =
& -(D_{m}q_{i})^{\dagger} D^{m}q^{i}
    -i\bar{\psi}_{q\dot{\alpha}}\bar{\sigma}^{m\dot{\alpha}\alpha}
    D_{m}\psi_{q\alpha}
    -i\psi_{\tilde{q}}^{\alpha}\sigma^{m}_{\alpha\dot{\alpha}}
         D_{m}\bar{\psi}_{\tilde{q}}^{\dot{\alpha}}
    \cr
& -\sqrt{2}\lambda^{\alpha i a}q_{i}^{\dagger}T_{a}\psi_{q\alpha}
    +\sqrt{2}\bar{\lambda}_{\dot{\alpha}}{}^{i a} q_{i}^{\dagger}T_{a}
      \bar{\psi}_{\tilde{q}}^{\dot{\alpha}}
    +\sqrt{2}\bar{\psi}_{q\dot{\alpha}}T_{a}
    q^{i}\bar{\lambda}^{\dot{\alpha}a}{}_{i}
    \cr
& + \sqrt{2}\psi_{\tilde{q}}^{\alpha}T_{a}q^{i}\lambda^{a}_{\alpha i}
   -\sqrt{2}\psi_{\tilde{q}}^{\alpha}T_{a}\psi_{q\alpha}B^{a}
   +\sqrt{2}\bar{B}^{a}\bar{\psi}_{q\dot{\alpha}}T_{a}
   \bar{\psi}_{\tilde{q}}^{\dot{\alpha}} \cr
& -q_{i}^{\dagger}T_{a}T_{b}q^{i}(B^{a}\bar{B}^{b}
  +B^{b}\bar{B}^{a})
 +\Fr{1}{2}(q^{i\dagger}T^a q^j + q^{j\dagger}T^a q^i)
   q_{i}^{\dagger}T_{a}q_{j}. \cr
}
}
The transformation rule for the hypermultiplet is
\eqn\aah{
\eqalign{
\delta q^{i}=
&-\sqrt{2}\xi^{\alpha i}\psi_{q\alpha}
+\sqrt{2}\bar{\xi}_{\dot{\alpha}}{}^{i}
\bar{\psi}_{\tilde{q}}^{\dot{\alpha}},
   \cr
\delta\psi_{q\alpha}=
&-\sqrt{2}i\sigma^{m}_{\alpha\dot{\alpha}}
   D_{m}q^{i}\bar{\xi}^{\dot{\alpha}}{}_{i}
   -2T_{a}q^{i}\bar{B}^{a}\xi_{\alpha i},
  \cr
\delta\bar{\psi}_{\tilde{q}}^{\dot{\alpha}}=
&-\sqrt{2} i\bar{\sigma}^{m\dot{\alpha}\alpha}D_{m}q^{i}\xi_{\alpha i}
 +2T_{a}q^{i}B^{a}\bar{\xi}^{\dot{\alpha}}{}_{i},
  \cr
}
}
and the transformation for the conjugate fields is
\eqn\aak{\eqalign{
\delta q_{i}^{\dagger}=
&-\sqrt{2}\bar{\psi}_{q \dot{\alpha}}
\bar{\xi}^{\dot{\alpha}}{}_{i}
   -\sqrt{2}\psi_{\tilde{q}}^{\alpha}\xi_{\alpha i},
   \cr
\delta \bar{\psi}_{q\dot{\alpha}}=
&\sqrt{2}i\xi^{\alpha i}D_{m}q_{i}^{\dagger}
\sigma^{m}_{\alpha\dot{\alpha}}
 -2 \bar{\xi}_{\dot{\alpha}}{}^{i}q_{i}^{\dagger}B^{a}T_{a} ,
 \cr
\delta\psi_{\tilde{q}}^{\alpha}=
&-\sqrt{2}i\bar{\xi}_{\dot{\alpha}}^{i}
     D_{m}q_{i}^{\dagger}\bar{\sigma}^{m\dot{\alpha}\alpha}
  -2\xi^{\alpha i}\bar{B}^{a}q_{i}^{\dagger}T_{a}.
  \cr
}
}

\newsec{Topological QCD}

Throughout this paper  we consider a simply connected
compact oriented smooth
Riemann four-manifold $X$ endowed with the Riemann metric
$g_{\m\n}$.  To be precise, we recall the mathematical set-up.
Let $G$ be a compact, connected simple Lie group with
Lie algebra $\eufm{g}$. We consider a principal $G$-bundle
$P$ over an oriented compact simply connected Riemann four-manifold
$X$. We denote $\adP$ for the bundle of the Lie algebras associated
to the adjoint representation. Picking an arbitrary (linear)
representation $R$ of $G$, we get a
vector bundle $E$ over $X$ associated with $P$ and the representation
$R$. We denote $\tilde E$ the conjugate (or dual)  vector bundle
of $E$ which is the associated bundle with $P$ and the  representation
$\tilde R$ conjugate to $R$. Then, the adjoint bundle
$\adP$ is a real sub-bundle of the endomorphism bundle
$End(E) = E\otimes \tilde E$.  If $G = SU(2)$, as an example,
$\adP$ consists of skew-adjoint trace-free endomorphisms.

\subsec{Twisting}

The twisting procedure of the minimal rigid $N=2$ SYM
 theory is explained very clearly in Witten's papers \WittenA\WittenB.
Since the supercharges transform, under the original
global symmetry group $H$, as the spinors, the global (rigid)
supersymmetric theory does not exist on the non-parallelizable space.
The twisting can be described as taking a diagonal
subgroup $SU(2)_{R^\pr}$ of $SU(2)_R\times SU(2)_I$ in $H$,
and regard
\eqn\bxa{
K^\pr = SU(2)_L\times SU(2)_{R^\pr},
}
as the rotation group instead of $K$. Under $K^\pr$,
the supercharges transform as $(1/2,1/2)\oplus (0,1)\oplus (0,0)$.
Then we take the $(0,0)$ component of the supercharge which transforms
as scalar, i.e., $Q = \vr^{\dot{\a}i}Q_{\dot{\a}i}$,
as the charge of global supersymmetry.
The resulting theory, called the TYM theory,  exists
on any oriented Riemann manifold.
The twisted transformation law is given by
\eqn\baa{\eqalign{
\delta A_\m &= i\vr \lambda_\m,\cr
\delta \lambda_\m &= -\vr D_\m \phi,\cr
\delta \phi &= 0,\cr
}\qquad
\eqalign{
\delta\chi_{\m\n} &= \vr H_{\m\n},\cr
\delta \bar\phi &= i\vr\eta,\cr
}\qquad\eqalign{
\delta H_{\m\n} &= i\vr[\phi,\chi_{\m\n}],\cr
\delta\eta &= \vr[\phi,\bar\phi],\cr
}
}
where the fields
$(A_\m,\l_\m,\phi, \chi_{\m\n}, H_{\m\n},\bar \phi, \eta)$
have the $U$-numbers $(0,1,2,-1,0,-2,-1)$.
Let  $T_\e(field)$ denote the variation of a field in
a gauge transformation generated by an infinitesimal parameter
$\e$. One finds
\eqn\bab{
(\delta_\vr\delta_{\vr^\pr} - \delta_{\vr^\pr}\delta_\vr)
(field) = T_\e(field),
}
where $\e^a = -2i\vr\vr^\pr\cdot \phi^a$.

In terms of the new global symmetry $K^\pr$,
the hypermultiplet transforms as follows;
\eqn\bac{\eqalign{
\psi_{q\a } & = (\Ha,0)^{-1}, \cr
\psi_{{\tilde q} \a} &= (\Ha,0)^{-1} ,\cr
q^{\dagger}&= (0,\Ha)^{0},\cr
\tilde q^{\dagger} & =(0,\Ha)^{0},\cr
}\qquad\eqalign{
\bar{\psi}_q^{\dot \a} &= (0,\Ha)^{+1}\,\cr
\bar{\psi}_{\tilde q}^{\dot \a} &= (0,\Ha)^{+1} ,\cr
q&= (0,\Ha)^{0},\cr
\tilde q & =(0,\Ha)^{0}.\cr
}
}
Note that all the complex bosons transform as the right-handed
spinor fields under $K^\pr$. The appearance of the spinor fields
after twisting
is the new feature of the TQCD.
At  first sight, one might conclude
that the TQCD exists only on spin manifolds.

To illuminate this point,
we should look at the twisting procedure more closely.
As explained in \WittenA\WittenB, the use of $K^\pr$, instead of
$K$ to generate rotations one should replace the standard stress
tensor $T$ by a modified stress tensor $T^\pr$. This may amount
to coupling the untwisted theory with external  gauge fields
by gauging $SU(2)_I$ and considering diagonal correlators
with the $SU(2)_I$ gauge fields related to the right-handed spin
connections $\o_R$.\foot{We used the argument in \CMR\ where
the twisting procedure as well as other approaches to the
cohomological
filed theories in general are explained in details.}
However, in the non-spin manifolds, the above
procedure is valid only on local region and can not be
defined consistently
at everywhere in the manifold.
Then, {\it it implies that the twisting in general is impossible
for a non-spin manifold even for the minimal theory.}
However, this is not true.
Instead we can use the $spin^c$ structure which exists
on arbitrary oriented Riemann $4$-manifolds. In fact,
the twisting procedure should involve  the $spin^c$
structure rather than the spin structure.\foot{
For an extensive review on the spin geometry, the readers
can consult
the excellent book \ML.}

The tangent space $TX$ of a compact oriented Riemann
$4$-manifold $X$
has the structure group $SO(4)$.  To define the spinor
 fields everywhere in $X$, one should
be able to lift $SO(4)$ to $Spin(4) = SU(2)\times SU(2)$.
The obstruction of defining a spin structure is measured by
the second Stifel-Whitney class $w_2(X) \in H^2(X;\BZ/2)$.
The spin structure does not exist unless $w_2(X) = 0$.
Instead, we consider any characteristic element $\eufm{c}$
such that
$\eufm{c}\in H^2(X;\BZ)$ and $\eufm{c} \equiv w_2(X)\hbox{ mod } 2$
which defines a $spin^c$ structure of $X$.  A $spin^c$ structure
defines  a pair of rank two hermitian vector bundle
$W^+_\eufm{c}$ and $W^-_\eufm{c}$
such that $det(W^\pm_\eufm{c}) = L^2_\eufm{c}$
where $c_1(L^2_\eufm{c}) =\eufm{c}$.\foot{We will
follow the physical notation of  Witten\WittenC.}
One can write
\eqn\pca{
W^+_\eufm{c} = S^+\otimes L_\eufm{c},
}
where $S^+$ and $L_\eufm{c}$ are both possibly
non-existing square roots
of some bundles. However, their tensor product exists everywhere.
A different integral lift of $w_2(X)\in H^2(X;\BZ/2)$ defines a different
$spin^c$ structure $\eufm{c}^\pr$ on $X$.
Changing  the $spin^c$ structure\foot{
Since we are considering the simply connected case, this is the
only way of changing the $spin^c$.}
on $X$ by an $c_1(\zeta) \in H^2(X;\BZ)$,
\eqn\pcaa{
\eufm{c}^\pr -\eufm{c} = 2c_1(\zeta)
}
amounts to twisting
the given $spin^c$ bundle by the associated  line bundle $\zeta$,
\eqn\pcb{
W^+_\eufm{c}= S^+\otimes L_\eufm{c} \rightarrow
W^+_\eufm{c}\otimes \zeta  = S^+\otimes L_\eufm{c} \otimes\zeta
=W^+_{\eufm{c}^\pr},
}
where $det(W^+_{\eufm{c}^\pr}) = L^2_{\eufm{c}^\pr}
= L^2_\eufm{c}\otimes \zeta^2$.
Of course $\eufm{c}^\pr$ is again a characteristic.
{\it Physically this amounts to say that one can have
many different choices
of the topological twisting.}

Now we can proceed the twisting procedure using
the $spin^c$-connection
instead of the spin connection. Here the $spin^c$-connection
means the usual spin connection as well as the connection of the
line bundle $L^2_\eufm{c}$ associated with a characteristic $\eufm{c}$.
The soul effect of the twisting using the $spin^c$ is that the
Dirac operator is twisted by the connection on $L^2_\eufm{c}$.
One may ask why the original TYM theory
does not refer to the $spin^c$ structure.  One obvious observation is
that there are no fields which transform as the spinor under $K^\pr$.
Later, it will become clear that the twisted minimal
theory (the original TYM) theory is actually independent of the choice
of the $spin^c$ structure. On the other hand,
the twisting of the $N=2$ hypermultiplet depends on the choice of
the $spin^c$. This fact endows  the TQCD with very rich
structures. Furthermore one should view that the original TYM theory
has a hidden symmetry generated by the variation of the $spin^c$
structure. This observation has a deeper implication.

To describe the topological transformation laws, one sets
\eqn\bad{
\xi^{\a i} =0,
}
and replace the internal indices $i,j,..$ of
$SU(2)_I$ by another $SU(2)_R$ index $\dot{\beta}$\foot{
This procedure leads to the so called {\it bispinor}.
In general one can regard the bispinor
as a special case of the spinor carrying an external index. This
point and related topics are very clearly and extensively discussed
in the chapters
$14$ and $15$ of ref.~\GSW.}
\eqn\bae{
\bar\xi^{\dot\a i} = \bar\xi^{\dot\a\dot\b} = -\e^{\dot\a\dot\b}\vr,
}
where $\vr$ is an anticommuting parameter.
{}From \aah\ and \aak\
we have\foot{Note that the untwisted bosonic field $q$ is a
$E$-valued scalar.
After the topological  twisting
that replaces $q^i$ by $q^{\dot\b}$,
$q$ becomes a section of
$(S^+\otimes L_\eufm{c})\otimes E$ for a $spin^c$ structure $\eufm{c}$.
In the mathematics literature, taking a certain tensor product to
an original bundle is usually referred as "twisting".  Thus,
it is legitimate to refer to the procedure as the twisting by  a
$spin^c$ structure.
}
\eqn\bag{
\eqalign{
\delta q^{\dot\a}=
 &-\vr \bar{\psi}_{\!\tilde{q}}^{\dot{\alpha}},
  \cr
\delta q_{\dot\a}^{\dagger}=
 &-\vr \bar{\psi}_{\!q \dot{\alpha}} ,
 \cr
}
\qquad\eqalign{
\delta\bar{\psi}_{\!\tilde{q}}^{\dot{\alpha}}=
& -i\vr \phi^{a}T_a q^{\dot{\a}},
  \cr
\delta \bar{\psi}_{\!q\dot{\alpha}}=
&i\vr q_{\dot\a}^{\dagger}\phi^{a}T_{a},
  \cr
}
}
where we set
\eqn\bba{\eqalign{
&B = \Fr{i}{2\sqrt{2}}\phi,  \cr
}\qquad\eqalign{
&\sqrt{2}\bar{\psi}_{\!\tilde{q}}^{\dot{\alpha}}
\rightarrow \bar{\psi}_{\!\tilde{q}}^{\dot{\alpha}},\cr
&\sqrt{2}\bar{\psi}_{\!q\dot{\alpha}}
\rightarrow \bar{\psi}_{\!q\dot{\alpha}},\cr
}
}
One finds
\eqn\bbb{
(\delta_\vr\delta_{\vr^\pr} - \delta_{\vr^\pr}\delta_\vr)
(field) = T_\e(field),
}
as expected.

On the other hand, the naive topological transformations
\eqn\bbc{\eqalign{
\delta \psi_{\!q\a}  \approx
 - i\vr \s^{\m}{}_{\a\dot\a} D_\m q^{\dot\a},\cr
\delta \psi_{\!\tilde q}^{\a}  \approx
 i\vr D_\m q_{\dot\a}^\dagger \s^{\m\dot\a\a},\cr
}
}
where we use the normalization,
\eqn\fgt{
\p_{q\a} \rightarrow \sqrt{2}\bar\p_{q\a},\qquad
\p_{\tilde{q}\a} \rightarrow \sqrt{2}\p_{\tilde{q}\a},
}
do not satisfy the relation \bbb\  in off-shell.
This is the typical situation and can be resolved by introducing
the auxiliary spinor fields $X_{q\a}$ and  $X^{\a}_{\tilde{q}}$ with $U=0$;
\eqn\bbd{
\eqalign{
\delta \psi_{\!q\a} &= -i\vr \s^{\m}{}_{\a\dot\a} D_\m q^{\dot\a}
                       +\vr X_{q\a},\cr
\delta \psi_{\!\tilde q}^{\a} &= i\vr D_\m q_{\dot\a}^\dagger
\bar \s^{\m\dot\a\a}
-\vr X_{\tilde{q}}^{\a}.\cr
}
}
Then one can find
\eqn\bbe{\eqalign{
\delta X_{q\a}
&=i\vr\phi^a T_a\psi_{\!q\a}
-i\vr \s^{\m}{}_{\a\dot\a}D_\m \bar\psi_{\!\tilde{q}}^{\dot\a}
+\vr \s^{\m}{}_{\a\dot\a}\lambda_\m^aT_a q^{\dot\a},
\cr
\delta X_{\tilde{q}}^{\a}
&=i\vr\psi_{\!\tilde{q}}^{\a}\phi^a T_a
-i\vr D_\m \bar\psi_{\!q\dot\a} \bar\s^{\m\dot\a\a}
+\vr q^\dagger_{\dot\a}\bar\s^{\m\dot\a\a} \lambda_\m^a T_a. \cr
}
}
One can check that the algebra is closed.

Before we leaves this section, we should emphasize that the
Dirac operator $\Fs{D}=\s^{\m} D_\m$ acts on
the $spin^c$ bundle twisted by $E$,
\eqn\dirac{
\s^{\m} D_\m: \G(W^+_\eufm{c}\otimes E)\rightarrow
\G(W^-_\eufm{c}\otimes E).
}
The square of the Dirac operator is given by the Weitzenb\"{o}ck formula
\eqn\wzbf{
\Fs{D}^*\Fs{D} = \nabla^*\nabla - F^+_A + \Fr{i}{2}p +\Fr{1}{4}R,
}
where $F^+_A$ is the self-dual part of the gauge field strength
and $p$ denotes Clifford multiplication by the curvature $2$-form
on $det(W^+_\eufm{c}) = L^2_\eufm{c}$ and
$R$ denotes the scalar curvature of the metric.

\subsec{Topological Action}

In general, the action functional of cohomological theory
can be written as a $Q$-commutator. To motivate
the correct formula we recall the familiar action functional
of TYM theory.  The topological action $S_{T}$  can be written
as
\eqn\caaa{
S_{T} = k/h^2 -i \{Q, V_{T}\}
}
where $k$ denotes the instanton number
\eqn\caaaa{
 k =  \Fr{1}{8\pi^2}\int \tr(F_A\wedge F_A),
}
and $V_T$ is given by
\eqn\caa{
V_{T} = \Fr{1}{h^2}\int d^4\!x\sqrt{g}\biggl[
\chi^{\m\n}_a\left(H^a_{\m\n} -i F^{+a}_{\m\n}\right)
-\Fr{1}{2} g^{\m\n}(D_\m\bar\phi)_a \l^a_{\n}
+\Fr{1}{8}[\phi,\bar\phi]_a \eta^a
\biggr].
}
We have
\eqn\caaa{\eqalign{
S_{T} = k/h^2+
 \Fr{1}{h^2}\int\! d^4\!x\sqrt{g}\biggl[
\left(H^{\m\n}_a - \Fr{i}{2}F^{+\m\n}_a\right)
\left(H_{\m\n}^a - \Fr{i}{2}F^{+a}_{\m\n}\right)
+\Fr{1}{4}F^{+\m\n}_a F^{+a}_{\m\n}
\cr
-i\chi^{\m\n}_a[\phi,\chi_{\m\n}]^a
+\chi^{\m\n}_a(d_A\l)^{+a}_{\m\n}
+\Fr{i}{2}g^{\m\n}(D_\m\eta)_a\l^a_\n
-\Fr{1}{2}g^{\m\n}(D_\m\bar\phi)_a(D_\n\phi)^a
\cr
-\Fr{i}{2}g^{\m\n}[\l_\m,\bar\phi]_a \l_\n^a
+\Fr{i}{8}[\phi,\eta]_a \eta^a
 +\Fr{1}{8}[\phi,\bar\phi]_a[\phi,\bar\phi]^a
\biggr].
\cr
}
}
A crucial property of the above action is that it is supersymmetric
in arbitrary orientable Riemann four-manifold. This looks obvious
but actually a highly nontrivial property since the above action directly
comes from the supersymmetric action defined on the flat space.
In general when verifying supersymmetry in curved space, one meets
the commutator of covariant derivative and the Riemann tensor
appears. However, the only appearance
of the commutator of covariant derivatives is on the fields which
transform as the scalars \WittenA. Note also  that there are no fields
which transform as the spinor  under $K^\pr$.
Thus,  neither the Riemann tensor  nor the curvature tensor associated
to the $spin^c$ structure appear. {\it Consequently, the TYM theory
is well defined for arbitrary oriented Riemann four-manifold
as well as completely independent of the $spin^c$ structure used
to define the twisting procedure.} That is, the $spin^c$ structure
is completely decoupled from the theory.

Some properties of the $V_{T}$ are
i) gauge invariance as well as invariance under $K^\pr$,
ii) all the fields are in the adjoint representation,
iii) the net $U$-number is $'-1'$,
iv) the non-degeneracy of the kinetic terms.

Actually, the above four properties almost completely
determine the possible form of the $V_{M}$
such that the total action functional can be written as
\eqn\cab{
S = S_{T} + S_{M} =k/h^2 -i\{Q, V_{T} + V_{M}\}= k/h^2 -i\{Q, V\}.
}
A useful choice of $V_{M}$ is
\eqn\cac{\eqalign{
V_{M} =
   \Fr{1}{h^2}\int d^4\!x\sqrt{g}\biggl[&
  -{i}\chi^{\m\n}_a q^\dagger\bar \s_{\m\n} T^a q
  +\left(X_{\!\tilde q}^{\a}\psi_{\!q\a}
      + \psi_{\!\tilde{q}}^{\a}X_{\!q\a}\right)
  \cr
&+i\left(q^\dagger_{\dot\a} \bar\phi_{a}T^a\bar\psi_{\!\tilde{q}}^{\dot\a}
  +\bar\psi_{\!q\dot\a} \bar\phi_{a}T^a q^{\dot\a}\right)
  \biggl].\cr
}
}
This form can be  easily seen from the Yukawa coupling of
the underlying theory.

We should remind the reader that $\c_{\m\n}$ is a self-dual tensor
field in the adjoint representation. Thus only the quantities
transforming as self-dual tensor and in the adjoint representation
can couple to $\c_{\m\n}$.
This becomes more transparent if we combine $V_T$ and $V_M$,
\eqn\unes{
V =  \Fr{1}{h^2}\int d^4\!x\sqrt{g}\biggl[
 \chi^{\m\n}_a \left[ H^a_{\m\n} - i\left(F^{+a}_{\m\n}
 +q^\dagger \bar\s_{\m\n} T^a q\right)\right] + \ldots.
 }
The above expression is completely sensible. Since
$q \in \G(W^+_\eufm{c}\otimes E)$ and
${q}^\dagger \in \G(\overline{W}^+_\eufm{c}\otimes \tilde{E})$,
the product $q\otimes  q^\dagger$ lies in
$$
W^+\otimes_\eufm{c} E\otimes \overline{W}^+_\eufm{c}\otimes \tilde{E}
\sim \O^0(End(E))\oplus\O^2_+(End(E))
$$ where $\O^0(End(E))$ and
$\O^2_+(End(E))$ denote the spaces of $End(E)$-valued zero-forms
and $End(E)$-valued self-dual-two-forms respectively.
Of course $F^+$ also lies in (the real subspace of) $\O^2_+(End(E))$.

After small computations, we have
\eqn\cad{\eqalign{
S_{M} =\Fr{1}{h^2}&\int\! d^4\!x\sqrt{g}\biggl[
-{i} H^{\m\n}_a q^\dagger\bar\s_{\m\n}T^a q
-2 X_{\!\tilde{q}}^{\a} X_{\!q\a}
+{i} X_{\!\tilde{q}}^{\a}\s^{\m}{}_{\a\dot\a}D_\m q^{\dot\a}
+{i}D_\m q^\dagger_{\dot\a}\bar\s^{\m\dot\a\a}X_{\!q\a}
\cr
&
-{i}\chi^{\m\n}_{a}\bar\psi_{\!q}\bar\s_{\m\n}T^a q
+{i}\chi^{\m\n}_{a}
  q^\dagger\bar\s_{\m\n}T^a \bar\psi_{\!\tilde{q}}
-{i}D_\m\bar\psi_{\!q\dot\a}\bar\s^{\m\dot\a\a}\psi_{\!q\a}
-{i}\psi_{\!\tilde{q}}^\a\s^{\m}{}_{\a\dot\a}
 D_\m\bar\psi_{\!\tilde{q}}^{\dot\a}
\phantom{\biggr]}\cr
&
+2 i\psi_{\!\tilde{q}}^\a\phi_a T^a \psi_{\!q\a}
-{2}i \bar\psi_{\!q\dot\a}\bar\phi_a T^a \bar\psi_{\!\tilde{q}}^{\dot\a}
+q^\dagger_{\dot\a}\l_{\m{a}} T^a \bar\s^{\m\dot\a\a}\psi_{\!q\a}
+\psi_{\!\tilde{q}}^\a \s^{\m}{}_{\a\dot\a}\l_{\m{a}} T^a
q^{\dot\a}. \phantom{\biggr]}\cr
&
+ q^\dagger_{\dot\a}\eta_a T^a \bar\psi_{\!\tilde q}^{\dot\a}
-\bar\psi_{\!q\dot\a}\eta_a T^a q^{\dot\a}
- q^\dagger_{\dot\a}T^aT^b\left(\phi_a\bar\phi_b +
 \phi_b\bar\phi_a\right)q^{\dot\a}
\biggr].\cr
}
}
The total action functional $S= S_T +S_M$ is given by the sum of
Eq.~\caaa\ and Eq.~\cad.  If one carefully compares the untwisted action,
(Eq.~\aaa +Eq.~\aag), with the twisted action $S$, one can see that both
do  in fact look identical at least for the flat Euclidean $4$-manifolds.
In addition to the gauge symmetry, the classical action functional
has the manifest $U$-number symmetry (the ghost number symmetry).
The situation concerning this $U$-number symmetry
is much like the one in the original TYM theory.

One can eliminate the auxiliary fields $H_{\m\n}$, $X_{\!q\a}$ and
$X_{\!\tilde{q}}^{\a}$ from the action functional.
The relevant terms in the action $S$ is
\eqn\cae{\eqalign{
S_0 = \Fr{1}{h^2}\int\!d^4\!x\sqrt{g}\biggl[
\left( H^{\m\n}_a
  - \Fr{i}{2}(F^{+\m\n}_a
  +q^\dagger\bar\s^{\m\n}T_a q
)\right)
\left( H_{\m\n}^a
- \Fr{i}{2}(F^{+a}_{\m\n}
+ q^\dagger\bar\s_{\m\n}T^a q
)\right)
\cr
+\Fr{1}{4}\left(F^{+\m\n}_a
+q^\dagger\bar\s^{\m\n}T_a q\right)
\left(F^{+a}_{\m\n}
+q^\dagger\bar\s_{\m\n}T^a q\right)
-{2} X_{\!\tilde{q}}^{\a} X_{\!q\a}
\cr
+{i} X_{\!\tilde{q}}^{\a}\s^{\m}{}_{\a\dot\a}D_\m q^{\dot\a}
+{i}D_\m q^\dagger_{\dot\a}\bar\s^{\m\dot\a\a}X_{\!q\a}
\biggr].\cr
}
}
One can eliminate $H_{\m\n}$ by setting
\eqn\caf{
H^{\m\n}_a =
   \Fr{i}{2}(F^{+\m\n}_a
   + q^\dagger\bar\s^{\m\n}T_a q)
}
We will use the notation that
\eqn\nooo{
s = F^{+\m\n}_a +q^\dagger\bar\s^{\m\n}T_a q.
}
The Gaussian integration over $X_{\!\tilde{q}}^{\a}$ and $X_{\!q\a}$
is equivalent to the squaring, $\Fr{1}{2}|k|^2$, where
\eqn\cag{
k = \s^\m D_\m q =\Fs{D}q.
}
Using the Weitzenb\"{o}ck formula \wzbf,
the action $S_0$ in \cae\ becomes\foot{The absence of the mixed
term, $\sim F^{\m\n}q^\dagger\bar\s_{\m\n}q$, actually originated
from the underlying untwisted theory,  (Eq.~\aaa +Eq.~\aag).
The new feature
of the twisted theory is the appearance of the terms proportional to
$p$ and $R$. These new data precisely indicate the property of the
topological twisting
of the hypermultiplets.}
\eqn\cah{\eqalign{
S_0
&= \Fr{1}{h^2}\int\!d^4\!x\sqrt{g}
\biggl[ \Fr{1}{4}|s|^2 +\Fr{1}{2}|k|^2\biggl]\cr
&= \Fr{1}{h^2}\int\!d^4\!x\sqrt{g}\biggl[
+\Fr{1}{4}F^{+\m\n}_a F^{+a}_{\m\n}
+ \Fr{1}{4}(q^\dagger\bar\s^{\m\n} T_a q)
                  (q^\dagger\bar\s_{\m\n} T^a q)
+\Fr{1}{2}g^{\m\n}D_\m q^\dagger_{\dot\a} D_\n q^{\dot\a} \cr
&\phantom{........................}
+\Fr{1}{8}R (q^\dagger_{\dot\a}q^{\dot\a})
+\Fr{i}{4}p_{\dot\a\dot\b}(q^{\dot\a\dagger}q^{\dot\b}
+q^{\dot\b\dagger}q^{\dot\a})
\biggl].\cr
}
}
The resulting action is $Q$ invariant after changing the transformation
law for $\chi_{\m\n}$ to
\eqn\caj{
\delta \chi^{a}_{\m\n} = \Fr{i}{2}\vr(  F^{+a}_{\m\n}
        +q^\dagger\bar\s_{\m\n}T^a q).
}
Similarly the eliminations of   $X_{\!\tilde{q}}^{\a}$ and $X_{\!q\a}$
give
\eqn\cak{
\eqalign{
\delta \psi_{\!q\a} &=- \Fr{i}{2}\vr \s^{\m}{}_{\a\dot\a}D_\m q^{\dot\a},
\cr
\delta \psi_{\!\tilde q}^{\a} &
= \Fr{i}{2}\vr D_\m q_{\dot\a}^\dagger\bar\s^{\m\dot\a\a} .
\cr
}
}
Thus the fixed point locus,
$\delta \chi_{\m\n} = \delta \psi_{\!q\a}
= \delta \psi_{\!\tilde q}^{\a} =0$,
of the theory is the space of solutions of
the following equations
\eqn\monopole{
\eqalign{
F^{+a}_{\m\n} +q^\dagger\bar\s_{\m\n}T^a q &=0,\cr
\s^{\m} D_\m q&=0,\cr
}
}
which is the non-Abelian version of the Seiberg-Witten monopole
equations. The solutions of the equation \monopole\
are of course the minimum energy
solutions of $S_0$.

Due to the fixed point theorem of Witten,
the path integral reduces to the fixed point locus.
Equivalently, in the semi-classical limit $h^2\rightarrow 0$
(which is exact)
the path integral receives the dominant contribution from the
minimum action configurations of $S_0$ which corresponds to
the solution space of the equation \monopole.
We refer the space of solution modulo
gauge symmetry as the SW moduli space. The moduli space of TQCD
can be specified by
\eqn\ccck{
\CM(X,G,R, k, \eufm{c})
}
where $X$ is the underlying oriented simply connected
Riemann $4$-manifold, $G$ is the gauge group, $R$
denotes the representation of $G$ carried by the hypermultiplets,
$k$ is the instanton number and $\eufm{c}$ is the $spin^c$ structure
on $X$ used to define the topological twisting.
If we consider the theory with no hypermultiplets, the moduli space
is the moduli space of ASD connections which will be denoted by
$\CM(X,G, k)$.

\subsec{The Classical Moduli Space}

Here the notion  of classical or quantum moduli space is used
in the sense of the Seiberg-Witten paper \SWb.
It is useful to collect the bosonic  terms of the action,
\eqn\bosonic{\eqalign{
S_{bose} = k/h^2
+ \Fr{1}{h^2}\int\!d^4\!x\sqrt{g}\biggl[
\Fr{1}{4}F^{+\m\n}_a F^{+a}_{\m\n}
+ \Fr{1}{4}(q^\dagger\bar\s^{\m\n} T_a q )
                  (q^\dagger\bar\s_{\m\n} T^a q)
+\Fr{1}{2}g^{\m\n}D_\m q^\dagger_{\dot\a} D_\n q^{\dot\a} \cr
+\Fr{1}{8}R (q^\dagger_{\dot\a}q^{\dot\a})
+\Fr{i}{4}p_{\dot\a\dot\b}(q^{\dot\a\dagger}q^{\dot\b}
 +q^{\dot\b\dagger}q^{\dot\a})
-\Fr{1}{2}g^{\m\n}(D_\m\bar\phi)_a(D_\n\phi)^a
\biggr].
\cr
}}
The theory has another important fixed point equations
\eqn\fixed{
\delta \l_\m = -\vr D_m \phi =0,\qquad
\delta\eta = \vr[\phi,\bar\phi] = 0.
}
The existence of a  solution $\phi$ of the above equation means
that the gauge connection is reducible.
Then the bundle reduces
to the direct sum of certain line bundles. Physically, it corresponds to
the flat direction of the superpotential and the gauge group $G$
spontaneously
breaks down to its maximal torus. As far as  concerning the
gauge symmetry,
the theory with one hypermultiplet has only two phases,
the unbroken phase
of  gauge symmetry and the broken phase to its maximal torus.

Furthermore the action functional
as well as the equation \monopole\ are invariant under
the global vector $U(1)$
symmetry (circled action)  generated by the scaling of spinor with
$e^{i\theta}$. Unlike the gauge symmetry which is a redundant
 symmetry
that should be fixed, this $U(1)$ symmetry is the true
symmetry of the theory.
This suggests that the minimum action configuration space is
degenerated.
Furthermore this $U(1)$ symmetry has the fixed points.
One fixed point is
obviously the configuration $q=0$.
The  solution space of \monopole\
or the minimum action configuration space is
the moduli space of $G$-ASD
connections modulo gauge symmetry.
The other fixed points are in
the Coulomb phase that a non-zero solution of $\phi$ of \fixed\
exist. If we consider the case that $G=SU(2)$ and $R$ is
the fundamental representation,
the gauge symmetry spontaneously breaks down to
$U(1)$ and the vector bundle $E$ reduces to
a sum of line bundles $E = \zeta_1\oplus\zeta_2$.
Then, for example,
$q= (q_1,q_2=0)$  is a fixed point since
the $U(1)$ action
on $q_{1}$ can be undone by the $U(1)$ gauge transformation.
Then the monopole equation \monopole\ becomes
the abelian monopole equation. We will return this point later
which turns out to be crucial.
If we consider the theory with more hypermultiplets,
this global symmetry will be enhanced accordingly.

In this  paper, we will not try to analyze the detailed
structure of the solution
space.
We   just would like to remark  on the  relation with the classical moduli
space of the underlying untwisted theory \SWa.
To establish an analogy, we consider
a $K3$ surface and the metric with vanishing scalar curvature.
Since $K3$ surfaces are spin manifolds,
we simplify the argument by choosing trivial
$spin^c$ structure for twisting, i.e. $\eufm{c}=0$.
Then the only possibility of the solution for \monopole\ is
just $q =0$,  $D_\m \phi = 0$ and $F^+ = 0$. Thus there are two
possibilities for the pairs $(q, A)$; i) $q=0$ and $A$ is an ASD
connection, ii) $q=0$ and $A$ is an abelian instanton (Coulomb Phase).
If we increase the number of hypermultiplets
 there can be other types of solutions which allow the certain
non-vanishing combination of spinors as a solution.
This immediately requires
the spinor to be a  covariant constant.  (Of course this is allowed
only for the
hyperK\"{a}hler manifold).
This branch is referred as  the Higgs branch.
What we tried to say is that the vanishing theorem
arguments are analogous to
studying the classical moduli space of the underlying physical theory.
But the relations between two seems to be not that close because of the
appearance of $R$ and $p$.
 Instead, we will demonstrate later that
the study of the minimum energy solution
space of TQCD is more closely
related to the quantum moduli space
of the underlying theory studied in
the second paper of Seiberg and Witten \SWb.

\subsec{The Fermionic Zero-Modes}

Now we discuss the geometric meaning of the fermionic zero-modes.
Here we refer to the fermionic zero-modes by the fields with
odd $U$-number (the ghost number) following the conventions
of the usual TYM theory.
The $\chi_{\m\n}$ equation of motion, modulo the gauge symmetry, gives
\eqn\daa{
(d_A\l)^{+a}_{\m\n}
-i\bar\psi_{\!q}\bar\s_{\m\n}T^a q
+iq^\dagger\bar\s_{\m\n}T^a \bar\psi_{\!\tilde{q}}
=0.
}
The $\psi_{\!\tilde{q}}^\a$ equation of motion, modulo the gauge symmetry,
gives
\eqn\dab{
i\s^{\m}{}_{\a\dot\a}D_\m\bar\psi_{\!\tilde{q}}^{\dot\a}
-\s^{\m}{}_{\a\dot\a}\l^a_{\m} T_a q^{\dot\a}
 =0.
}
These two equations \daa\ and \dab\ are  precisely the linearization of
the equations \monopole\
by identifying the infinitesimal variations
$ (\delta A, \delta q,\delta q^\dagger)$ with
$(\l,i\bar\psi_{\!\tilde{q}}, i\bar\psi_{\!{q}})$.
Note that all $\l$,  $\bar\psi_{\!\tilde{q}}$ and
$\bar\psi_{\!{q}}$ have the $U$-number (the ghost number)
$1$.
Furthermore, the $\eta$ equation of motion gives
\eqn\dac{
D_\m \l_{a}^\m
+iq^\dagger_{\dot\a} T_a \bar\psi_{\!\tilde q}^{\dot\a}
+i\bar\psi_{\!q\dot\a} T_a q^{\dot\a} = 0,
}
which expresses the fact that
$(\l,i\bar\psi_{\!\tilde{q}}, i\bar\psi_{\!{q}})
= (\delta A, \delta q,\delta q^\dagger)$
is orthogonal to the gauge directions.
Thus the zero-modes of  $(\l, \bar\psi_{\!\tilde{q}},\bar\psi_{\!{q}})$
are the tangent vectors of $\CM$.
In the generic situation, the number of the zero-modes
$(\l, \bar\psi_{\!\tilde{q}},\bar\psi_{\!{q}})$ corresponds to
the real dimension of the moduli space $\CM$. Of course,
the moduli space is the moduli space of ASD connections
in the absence of the hypermultiplets.

The dimension of the moduli space can be easily calculated by
applying the Atiyah-Singer index theorem.
After twisting the complex boson
$q$  ($\tilde q$) can be viewed as  a section of
$W^+\otimes E$ ($W^+\otimes \tilde E$).
By defining $\delta_A = d_A^+ \oplus d^*_A$,
the above equations \daa,
\dab\ and \dac\ can be summarized as
\eqn\daaf{
\delta_A \oplus \Fs{D} : \O^1(\adP)\oplus
(W^+_\eufm{c} \otimes E) \rightarrow
\O^0(\adP) \oplus \O^2_+(\adP)\oplus(W^-_\eufm{c}\otimes E).
}
Thus the real virtual dimension is given by the sum of the index of
$\delta_A$ and the twice of the usual Dirac index of $\Fs{D}$.
The index of $\delta_A$ is well known, and  gives the virtual
dimension  of the moduli space $\CM(X,G,k)$ of anti-self-dual
connections,
\eqn\daag{\eqalign{
index(\delta_A)
&= dim (\CM(X,G,k))  = p_1(\adP^\BC) -\Fr{dim(G)}{2}(\chi + \s)\cr
&= 4 c_2(adj)k(P) -\Fr{dim(G)}{2}(\chi + \s),\cr
}}
where $\chi$ and $\s$ denote the Euler characteristic
and the signature of $X$,
and $k(P)$ denotes the instanton number
\eqn\daai{
k(P) =\Fr{1}{8\pi^2}\int_X\tr F_A\wedge F_A.
}
The index  of the twisted Dirac operator associated to a
$spin^c$ structure $\eufm{c}$ is given by
\eqn\daaj{
index(\Fs{D}) = \int_X e^{\Fr{1}{2}c_1(L^2_\eufm{c})}
ch(E)\wedge \hat{A}(X),
}
where
the trace inside $ch(E)$  should be
taken in the representation $R$ which the hypermultiplet carries.
The rank of $E$ is identical to the dimension of the
representation $rk(E) =dim (R)$.
We have\foot{
Note that
$
T^a T_a  = -C_2(R) I,\quad
Tr_R T^a T^b = -T(R)\delta^{a b},
\quad
c_2(R) = T(R)\Fr{dim (R)}{dim (adj)}.
$
}
\eqn\daal{\eqalign{
index(\Fs{D})
&= \Fr{1}{2}\int_X\!\!\left(c_1^2(E) - 2 c_2(E)\right)
+ \Fr{rk(E)}{2}c_1^2(L_\eufm{c})
+\int_X\!\! c_1(L_\eufm{c})\wedge c_1(E)
-\Fr{rk(E)}{24}\int_X p_1(X)\cr
&= -2T(R)k(P) +\Fr{rk(E)}{2}c_1^2(L_\eufm{c})
+\int_X\!\! c_1(L_\eufm{c})\wedge c_1(E)-\Fr{rk(E)}{8}\s.\cr
}
}
Putting everything together we have
the real virtual dimension of the moduli space
\eqn\daam{\eqalign{
dim \CM(X,G,R,k,\eufm{c}) = &dim \CM(X,G,k) + 2\, index(\Fs{D})\cr
=& 4\left(c_2(adj) - T(R)\right)k(P) + dim (R)c_1^2(L_\eufm{c})
+2\int_X c_1(L_\eufm{c})\wedge c_1(E)
\cr
&-\Fr{dim(G)}{2}(\chi
+\s)-\Fr{dim(R)}{4} \s.
 \cr
 }
 }

\newsec{The Observables and the Correlation Functions}

\subsec{The Observables}

The partition function or the correlation functions of topological
observables in TYM theory are the differential topological invariants
(the Donaldson invariants) of the smooth four-manifold.
It is straightforward to show that the corresponding quantities
in TQCD are also differential topological invariants.
For simplicity we consider the simply connected four-manifolds.
We also restrict our attention to the the $SU(2)$ case with
hypermultiplet in the fundamental representation. The generalization
for the other groups and representations should be straightforward.

The topological observables of the original TYM theory is given
by
\eqn\eaa{
\eqalign{
\Theta(x) &= \Fr{1}{8\pi^2}\tr \phi(x)^2,\cr
O_\S & = \Fr{1}{4\pi^2}\int_\S \tr\left(i\phi F
+\Fr{1}{2}\l\wedge\l\right),\cr
}
}
where $\S \in H_2(X;\BZ)$. The geometrical meaning of the
observables is well-known.
The observables $\Theta$ and $\O_\S$ also serve as
topological observables in the TQCD.

Since there are no additional fields, coming from the hypermultiplet,
which is $Q$ invariant, one may think that there is no new
additional topological observables. However, one can find
that the combination,
\eqn\eab{
i\phi^a  q^{\dagger}_{\dot\a}T_a q^{\dot\a}
+ \bar\psi_{\!q\dot\a}\bar\psi_{\!\tilde{q}}^{\dot\a},
}
is $Q$-invariant. Note that the above combination carries
the ghost number $2$ similar to the observables $O_\S$.
Note that the two-form
$\tr\left(i\phi F +\Fr{1}{2}\l\wedge\l\right)$
is $Q$-invariant up to an exact term which leads
to the fact that $O_\S$ is $Q$-invariant and depends
only on the homology class of $\S \in H_2(X;\BZ)$.
However the above combination is absolutely $Q$-invariant and
transforms as a scalar similar to the observables $\Theta$.
One can change the combination to transform as
a vector, a second rank tensor, etc.
In any case, the above observable does
not depend on $H_2(X;\BZ)$ at all and trivial in the
$Q$-cohomology. In fact, the term \eab\
is a part of  the bare mass term of the hypermultiplet.

\subsec{The Correlation Functions}

The partition function of the theory is formally defined as
\eqn\partition{
Z = \Fr{1}{\hbox{vol}(\CG)}\int \CD Y\exp ( - S),
}
where $\CG$ is the group of the gauge transformations
and $\CD Y$ denotes the path integral measure.
We will consider only the generic case that there are
only fermionic zero-modes of
 $(\l,i\bar\psi_{\!\tilde{q}}, i\bar\psi_{\!{q}})$
which correspond to the tangent vectors of the moduli space.
Though the action $S$ has the global $U$-number symmetry at
the classical level, the path integral measure is not invariant
under $U$ due to the zero-modes of
$(\l,i\bar\psi_{\!\tilde{q}}, i\bar\psi_{\!{q}})$ which carry $U=1$.
The net violation of $U$, due to the zero-modes, is identical to the
 dimension
$dim \CM(X,G,R_i,k,\eufm{c})$ of the moduli space.
The partition function can be non-zero if the dimension of the
moduli space is zero. By the standard analysis of any
cohomological
theory, the partition function reduces to an algebraic sum of
the points of the moduli space up to sign.\foot{Combining the
arguments of \DonaldsonB\ and \WittenC, it is straightforward to
define the orientation of the moduli space.}

For the generic choice of the metric, the moduli space
is a smooth manifold with actual dimension equals to
the index.
Let the dimension of the moduli space be  even
$dim  \CM(X,G,R_i,k,\eufm{c}) = 2d$.
Then the correlation functions
\eqn\corre{\eqalign{
\left< \Theta^r \prod_{i=1}^{s} O_{\S_i}\right>
&=  \Fr{1}{\hbox{vol}(\CG)}\int \CD Y\; e^{- S}\cdot
\Theta^r \prod_{i=1}^{s} O_{\S_i} \cr
& = e^{-k/h^2}\times \Fr{1}{\hbox{vol}(\CG)}\int \CD Y\; e^{i\{Q,V\}}
\Theta^r \prod_{i=1}^{s} O_{\S_i},\cr
}
}
can be non-vanishing if and only if
\eqn\selection{
d = 2r + s,
}
due to the ghost number anomaly.
It is straightforward to show that the above correlation
functions and the partition function are metric independent
at least for manifolds with $b_2^+ > 1$.
It is also straightforward to show that the quantity
\eqn\ccor{
\Fr{1}{\hbox{vol}(\CG)}\int \CD Y\; e^{i\{Q,V\}}
\Theta^r \prod_{i=1}^{s} O_{\S_i}
= \int_\CM \widehat\Theta\wedge ...\wedge
\widehat\Theta\wedge\widehat{O}_{\S_1}...
\wedge \widehat{O}_{\S_1},
}
reduces to the integration of the cup-products of
$\widehat\Theta \in H^4(\CM;\BZ)$
and $\widehat O_{\S_i} \in H^2(\CM;\BZ)$ where
$\widehat\Theta$ and
$\widehat O_{\S_i}$ are just the restrictions and the reductions of
$\Theta$ and $O_{\S_i}$. We will denote the above invariants
as $\hbox{DSW}_{X,G, R,k,\eufm{c}}(\S_1,...,\S_s, (pt)^{d(k)-2s})$.

\newsec{Some Prospectives}

\subsec{On the Seiberg-Witten invariants}

We consider twisted $N=2$
super-Maxwell theory with one hypermultiplet.
To do the topological twisting
we should choose a $spin^c$ structure $\eufm{c}$.
Then the monopole is a section of $W^+_\eufm{c} \otimes E$ where
$E$ is a line bundle whose curvature is the gauge field
strength of the $U(1)$ gauge symmetry.
According  to  the famous Dirac quantization condition,
the first Chern class $c_1(E)$ corresponds to the charge of
 the magnetic monopole.
The  formula derived by Seiberg-Witten
precisely quantize the monopole charge to be
an integral value \SWa.
That is, $c_1(E) \in H^2(X;\BZ)$. As we reviewed before,
taking the tensor products
$E$ to $W_\eufm{c}^+$ amounts to changing the $spin^c$ structure
by $c_1(E)$.
Thus, both $W^+_\eufm{c}$ and  $W^+_\eufm{c} \otimes E$
are $spin^c$ bundles associated to the different choices of the
$spin^c$ structure on $X$.  In other words, they correspond to
the different choice of the topological twisting.
Then, it is unnecessary to specify the
$U(1)$-bundle $E$ associated with the gauge group,
as Witten did.
Once all these are understood, our formalism  precisely
recovers the  original Seiberg-Witten monopole
equation \WittenC\ and the corresponding topological
field theory for each choice of the $spin^c$ structure
or the topological twisting.\foot{Precisely speaking, one
obtains a perturbed version of the monopole equation,
however, one can removes the perturbation.}

\subsec{On a $N=4$  theory}

Now we consider the case when the  hypermultiplet is in the adjoint
representation. The underlying physical theory corresponds to
the $N=4$ SYM theory. Interestingly enough,
the dimension
of the moduli space is independent of the instanton numbers,
\eqn\ppa{\eqalign{
dim( \CM)_{\beta =0}
=& dim (adj) \,c_1^2(L_\eufm{c})
 +2\int_X c_1(L_\eufm{c})\wedge c_1(E)
 \cr
 &-\Fr{dim(G)}{2}(\chi +\s)
- \Fr{dim (adj)}{4}\s.
\cr
}}
Assume that  the dimension is a positive even integer $2d$.
Since the theory has vanishing $\beta$-function,
we have natural  a theta
angle $\theta$. We turn on the theta term in
the action,
\eqn\ppb{
S \rightarrow S +\Fr{i \theta}{2\pi} k,
}
and  introduce
\eqn\aas{
\tau = \Fr{\theta}{\pi} + \Fr{8\pi i}{h^2}, \qquad
q = e^{2\pi i \tau}.
}
Since the   dimension of the moduli space is independent of
the instanton number, one can naturally sum up
the correlation function
\eqn\paat{\eqalign{
\sum_k \left< \prod_{i=1}^{d} O^\pr_{\S_i}\right>_k
&= \sum_k\left( \Fr{1}{\hbox{vol}(\CG)}\int \CD Y\;
e^{ 2\pi \tau k +i\{ Q, V\}}
\prod_{i=1}^{d} O^\pr_{\S_i}\right)\cr
&= \sum_k q^k \hbox{DSW}(\S_1,...\S_d)_k.\cr
}}
According to the Montonen  -Olive duality conjecture\MO,
the refinements of Vafa-Witten\VW\ and
Seiberg-Witten\SWb,
one can expect that the above expression is  modular covariant.
If we consider a manifold with $b^+ =1$, following the
suggestion of Vafa and Witten \VW, one can also expect that
there will be
holomorphic anomaly.

At least for the one case we can confirm the expectation.
Consider the $K3$ surfaces. A $K3$ surface is simply connected
oriented spin manifold with trivial canonical line bundle.
Since it has a spin structure, we can choose
$c_1(L^2_\eufm{c}) =\eufm{c} = 0$.
Now we
choose $G=SU(2)$, $N_f =1$ and $R=adj.$. Then we have
$dim(\CM) = 0$.  Thus we have a well-defined partition function
$Z_k$ and
\eqn\uiop{
\sum_{k} q^k Z_k.
}
The underlying theory is in fact the $N=4$ super-Yang-Mills
theory. There are three different ways of twisting and two of them
were studied by Yamaron\foot{
One can easily check that the remaining one possibility
corresponds to  our case.
Note also that the other two different ways of twisting
are independent of the $spin^c$ structure.
}\Yamaron\ and one of the two was studied by Vafa and Witten\VW.
They also showed that the resulting theory
defines a modular covariant form.
Since the canonical line bundle of a $K3$ surface is trivial,
twisting does nothing.
Thus, the above expression \uiop\ should be also the
modular covariant form.

\subsec{On the cobordism and the electro-magnetic duality}

Throughout this paper, we examined  formal properties of
the TQCD. In this section, we  demonstrate that the TQCD
can be a powerful tool to classify the N=2 supersymmetric QCD
up to the electro-magnetic duality.
For the $G=SU(2)$ and  $N_f = 0,1,2,3,4$ numbers of
hypermultiplets all in the fundamental representations,
this problem has been completely solved by Seiberg and Witten\SWa\SWb.
Using the similar analysis, the $G=SU(N)$ and $N_f=0$ case was
studied in  \AF\KLYT.

One of the possible strategy of finding the dual theory is to use
the twisted versions of the various possible theories and
to compare the differential-topological answers.
This method was successfully used to test the $S$-duality of the $N=4$
SYM theory \VW.  For example,
if one studies the twisted $N=2$ super-Maxwell theory
with one hypermultiplet, one would be able to see the great similarity
between the basic class and the Seiberg-Witten class. However,
the precise relation of the Donaldson invariants and the Seiberg-Witten
invariants can not be obtained by just looking at the  twisted $N=2$
super-Maxwell theory with one hypermultiplet. Furthermore,
the above strategy is rather difficult and unilluminating.

Actually, one can do much better. It turns
out that the TQCD with  one hypermultiplet understands the strong
and the weak coupling limits of the underlying physical $N=2$
supersymmetric Yang-Mills theory  coupled with no matter.
To be concrete, we consider the $N=2$ SYM theory with gauge group $SU(2)$.
The underlying mathematical argument of the following  procedure is
due to Pidstrigach \PG, and Tyurin \TyurinB\ and  Mrowka \MR.

One can twist the theory to get the standard TYM theory.
One consider
the exact  semi-classical limit of the theory and the path integral
has the dominant
contribution from the moduli space of $SU(2)$ ASD connections.
Our goal is to find the dual theory of the underlying physical theory.
One can assume that the dual theory  also has
a $N=2$ supersymmetric theory.
One can twist the dual
theory and this twisted theory is essentially characterized
 by the semi-classical
data of the untwisted theory. The duality means that those data talk about
the strong-coupling limit of the original $SU(2)$, $N=2$ SYM theory.
Of course this gives the dual description of the $SU(2)$ TYM theory.

Now we consider the $N=2$ SYM theory coupled with one hypermultiplet
$(N_f =1)$ carrying the fundamental representation of $SU(2)$.
Picking a $spin^c$ structure $\eufm{c}$, one can twist
the theory as we discussed.
We consider the exact semi-classical limit
of the twisted theory that the path integral is
localized to the fixed points
locus \caj\cak\ which is the solution space of the equations \monopole.
The formal dimension of the moduli space, the space of solution modulo
gauge symmetry is then
\eqn\mma{
dim\; \CM(k,R,\eufm{c}) = 8k -\Fr{3}{2}(\chi + \s)
+ \Fr{1}{2}\eufm{c}^2 -2k -\Fr{1}{2}\s.
}
On the other hand the dimension of the ASD connections is
given by
\eqn\mmaa{
dim\; \CM(k) = 8k -\Fr{3}{2}(\chi + \s).
}
Thus the codimension of $\CM(k)$ in $\CM(k,R,\eufm{c})$ is
 \eqn\iaa{
dim\; \CM(k,R,\eufm{c}) - dim\; \CM(k) =\Fr{1}{2}\eufm{c}^2 -2k
 -\Fr{1}{2}\s.
 }
As we vary the instanton number $k$ the codimension also vary.
However, one can choose the different  topological twisting
for the hypermultiplet such that,
\eqn\iab{
\Fr{1}{2}\eufm{c}^2 -2k -\Fr{1}{2}\s = 2,
}
the moduli space $\CM(k,R,\eufm{c})$ is always two dimension higher
than the moduli space $\CM(k)$ of ASD connections.

Recall the the TQCD with one hypermultiplet has a global $S^1$ action
on $q$. So
one can form the quotient space
\eqn\cccka{
\CM(X,G,R, k, \eufm{c})/U(1),
}
which is one dimension higher than $\CM(k)$.
The $S^1$ action has an obvious fixed point $q= 0$ which is
the moduli space $\CM(k)$ of ASD  connections.
The other fixed points is in the Coulomb phase
in which the connection $A$ is reducible, $E = \zeta_1\oplus \zeta_2$,
and $A$ preserves the splitting.
Now spinor $q$ has the two components $q =(q_1,q_2)$.
If $q_2=0$ then $(A,q)$ is a fixed point. $A$ induces a connection
$B$ in $\zeta$ and the pair $(B, q_1)$ solves the (perturbed)
$U(1)$ monopole equation.  More precisely,  one can obtain
the Seiberg-Witten monopole equation for
$W^+_\eufm{c}\otimes \zeta$, i.e.,
$det(W^+_\eufm{c}\otimes \zeta) = L^2_\eufm{c} \otimes \zeta^2$.

Then, the quotient $\CM /S^1$  looks like a cobordism
between $\CM_{ASD}$ and the cone on projective spaces corresponding
to the reducible solutions. Thus the evaluation of the Donaldson
map on the moduli space $\CM_{ASD}$ is the same as the evaluation
on the projective space. Thus the $U(1)$ Seiberg-Witten monopole
theory can be an alternative description of the Donaldson theory.

The above observation has  profound physical implications.
Note that the topological interpretation of the
path integral of the TQCD is essentially
based on the weakly coupled
ultraviolet limit of the underlying untwisted  $N=2$ supersymmetric QCD.
However, the electro-magnetic duality of the $N=2$ super-Yang-Mills
theory without matter is a genuine quantum theoretical effect.
The above example shows that  the dual theory of the $N=2$ super-Yang-Mills
theory may be determined by
studying  purely semi-classical properties of  TQCD.
Of course, the above duality
is for the twisted theory. However, since the underlying $N=2$
super-Yang-Mills
theory is asymptotically free its dual theory is a effective theory of
 strongly coupled infrared limit.
In the large scaling limit, the Riemann manifold looks
much like a flat Euclidean manifold. Thus the underlying theory for the
dual description of the Donaldson theory can be viewed as the dual theory
of the $N=2$ super-Yang-Mills theory. This is the reverse of the way
Witten arrived in his monopole invariants from the  dual theory
of the physical $N=2$ super-Yang-Mills theory.

The TQCD provides much more general picture than the cobordism
argument. If we consider general groups and representations
it is generically impossible to fix the codimension to $2$.
Then, the quotient space like \cccka\ does not lead to
the cobordism mentioned above.  However,
it is always possible to express the Donaldson invariants
in terms of the Seiberg-Witten invariants.
In a forthcoming paper \HPPb, we will present a explicit and general relation
between the  Donaldson invariants and the
Seiberg-Witten invariants based on  the twisted $N=2$,
super-Yang-Mills theory coupled with hypermultiplet having the bare
mass.

\subsec{The summary and generalization}

For a compact connected simple Lie group $G$ with
the Lie algebra $\eufm{g}$,
we consider  an arbitrary  sequence of representations
$(R_1,R_2,\ldots, R_{N_{f_{\!c}}})$
of $G$ such that
\eqn\jaa{
a(R_1,R_2,...,R_{N_{f_{\!c}}})
=  \left(c_2(adj) - \sum_{i=1}^{N_{f_{\!c}}} T(R_i)\right) = 0.
}
Throughout this section, we will use the fixed sequence
of representations given above.

Now we consider the sequence of the $N=2$ super-Yang-Mills
theories on the flat Euclidean $4$-manifold  $\BR^4$
coupled with $N_f$ number of hypermultiplets carrying
the representations  $R_1,R_2,\ldots, R_{N_f}$
where $N_f \leq N_{f_{\!c}}$.
The $N=2$ super-Yang-Mills theory,
in terms of the $N=1$ superspace language,
consists of a vector multiplet $W_\a$ and a chiral multiplet
$\Phi$ in the adjoint representation.  The theories with $N_f\neq 0$
have additional $N_f$ pairs of conjugate chiral multiplets $Q_i$
and $\tilde Q_i$ in the representations $R_i$ and $\tilde R_i$,
respectively. The one-loop $\beta$-function of the
theory is given by
\eqn\jab{
\beta(h) = -\Fr{h^3}{(4\pi)^2} a(R_1,R_2,...,R_{N_{\!f}}).
}
Thus we have the sequence of the asymptotically free theories
and the critical theory $N_f = N_{f_{\!c}}$ with
vanishing $\beta$-functions\foot{
The $\beta$-function \jab\ is exact in the all loop perturbations.
There are non-perturbative instanton corrections
which was studied in \SBa.
However, it was argued that the $\beta$-function of a critical theory
vanishes exactly \SWb.
}
\eqn\jac{
SY\!M\equiv SQC\!D(0)
\leftrightarrow SQC\!D(1)
\leftrightarrow  SQC\!D(2)
\leftrightarrow  \ldots  SQC\!D(N_{f_{\!c}-1})
\leftrightarrow  SQC\!D(N_{f_{\!c}}).
}
Associated to this, we have a sequence of inequalities
\eqn\jad{
a(0) > a(R_1) > a(R_1, R_2)> \ldots
> a(R_1,R_2,\ldots, R_{N_{f_{\!c}}-1})
>  a(R_1,R_2,\ldots, R_{N_{f_{\!c}}}) = 0.
}

Now we consider the compact oriented simply connected
Riemann
manifold $X$.  A $spin^c$ structure $\eufm{c} \in H^2(X;\BZ)$
is an integral lift of the Stifel-Whitney class
$w_2(X) \in H^2(X;\BZ/2)$, i.e.
$\eufm{c} \equiv w_2(X) \hbox{ mod } 2$.
The space $H^2(X;\BR)$ of
harmonic two-forms on $X$ is an
$b_2$-dimensional flat  space
with signature  $(b_2^+ - b_2^-)$. The space $H^2(X;\BZ)$ is
the integral lattice in $H^2(X;\BR)$. Then, the set $H^2_s(X;\BZ)$
of all $spin^c$ structure is an affine sublattice of $H^2(X;\BZ)$.
Obviously, there are $b_2$ independent generators $\eufm{s}$ of
the transitive action on $S$.
We consider a vector bundle $P$ on
 $X$ with the reduction of the structure group to $G$.
We denote $\adP$ for the bundle of the Lie algebras associated to
the adjoint representation.
For the sequence $\{R_i\}$ of the representations
we have a sequence of vector bundles $\{E_i\}$ over $X$
which are associated with $V$ and the sequence $\{ R_i\}$.
We denote $\{ \tilde E_i\}$ the sequence
of the conjugate (or dual)  vector bundles of $E_i$.
The adjoint bundle $\adP$ is a real sub-bundle of the
endomorphism bundles
$End(E_i) = E_i\otimes  \tilde{E}_i$.

To define the topological
twist of the $N=2$ supersymmetric QCD with $N_f$ hypermultiplets,
one should specify  $N_f$ independent $spin^c$ structures
$\eufm{c}_1, \eufm{c}_2, \ldots, \eufm{c}_{N_f}$.
Then we twist the sequence of the hypermultiplets
with corresponding sequence of the $spin^c$ structures.
This amounts to twisting the vector bundle $E_i$
by the $spin^c$ bundle $W^+_{\eufm{c}_i}$ in the sense that
the vector bundle $E_i$-valued complex boson $q_i$ in the
$i$-th hypermultiplet should be regarded as a
section of $ W^+_{\eufm{c}_i}\otimes E_i$,
\eqn\jae{
q_i \in \G( W^+_{\eufm{c}_i}\otimes E_i).
}
We call the resulting topological QCD denoted by
$T(\eufm{c}_1, \eufm{c}_2, \ldots, \eufm{c}_{N_f})$
as the TQCD associated to the sequence
$\eufm{c}_1, \eufm{c}_2, \ldots, \eufm{c}_{N_f}$
of the $spin^c$ structures.
We call a TQCD, $T(\eufm{c}_1, \eufm{c}_2, \ldots, \eufm{c}_{N_f})$,
critical if $N_f = N_{f_{\!c}}$ and we call a TQCD, $T_0$, minimal
if $N_f =0$.

We can vary each $spin^c$ structures  independently to define
another TQCD. In
general we have different TQCD
for the different choice of the $spin^c$ structures.
The minimal theory $T_0$ is
independent of the choice of the $spin^c$ structure.
The TQCD with $N_f$ hypermultiplets has
$N_f$ independent choices of the
$spin^c$ structure and in each $spin^c$ structure
has $b_2$ independent ways of variations.
Thus the space of the TQCD with $N_f$ hypermultiplets
is $b_2\cdot N_f$ dimensional, i.e.,
$H^2_s(X;\BZ)\otimes \ldots\otimes H^2_s(X;\BZ)$.

After twisting, we get a topological action which has a
global supersymmetry.
According to  the fixed point theorem of Witten,
the path integral reduces
to an integration over the fixed point locus.
The fixed point equations of the theory are
\eqn\mono{\eqalign{
F^+_{\m\n}  +  q^\dagger_i \bar\s_{\m\n}q^i = 0,\cr
\s^\m D_\m q^i = 0,\cr
}}
where $q_i \in \G(W^+_\eufm{c_i}\otimes E_i)$,
$F^+_{\m\n} \in \O^2_+(\adP) $
is the self-dual part of the curvature tensor on $P$ and
\eqn\dirac{
\s^\m D_\m :  \G(W^+_\eufm{c_i}\otimes E_i)\longrightarrow
\G(W^-_\eufm{c_i}\otimes E_i)
}
is the twisted Dirac operator of the $spin^c$ structure.
We denote the space of all solutions $(A, q^i)$ modulo gauge
symmetry by
\eqn\moduli{
\CM(k,\eufm{c_1},\ldots,\eufm{c}_{N_f})
}
where $k$ denote the instanton number. Note that
$\CM(k)$ is the moduli space ASD connections.
The path integral of a theory $T(\eufm{c_1},\ldots,\eufm{c}_{N_f})$
then localized to $\CM(k,\eufm{c_1},\ldots,\eufm{c}_{N_f})$ modulo
gauge symmetry.  The theory
$T(\eufm{c_1},\ldots,\eufm{c}_{N_f})$
is a twisted version of the underlying asymptotically free
$SQC\!D(N_f)$.
Thus, the cohomological description essentially rely on the ultra-violet
weak coupling limit of  $SQC\!D(N_f)$. We can interpret the moduli
space $\CM(k,\eufm{c_1},\ldots,\eufm{c}_{N_f},\eufm{c}_{N_f} )$
characterizes the weak coupling limit of  $SQC\!D(N_f)$.
More precisely, we should consider every moduli spaces
$\CM(k,\eufm{c_1},\ldots,\eufm{c}_{N_f})$
by varying the $spin^c$ structures to know about the
weak coupling limit of $SQC\!D(N_f)$.
The virtual dimension of the moduli space
can be easily calculated. For
$c_1(P)=0$, we have\foot{Of course we should only consider
the  $spin^c$ structures such that the dimension of
the moduli space is non-negative.}
\eqn\daam{\eqalign{
\hbox{dim}\CM&(k,\eufm{c_1},\ldots,\eufm{c}_{N_f})\cr
&
=4\, a(R_1,R_2,...,R_{N_{f}})\;k
 -\Fr{\hbox{dim}(G)}{2}(\chi +\s)
 + \Fr{1}{4}\sum_{i=1}^{N_f}\hbox{dim}(R_i)(\eufm{c_i}^2-\s)
 \cr}
}
The theory has other important fixed point equations,
\eqn\reduction{
D_\m \phi = 0,\qquad [\phi,\bar\phi] = 0.
}
If there are non-zero solution of $\phi$,
the connection is reducible.
Then the group $G$ spontaneously break down
to its maximal torus.

The theory has additional global symmetry acting
on hypermultiplets.
In general  the global internal symmetry  always contains
\eqn\glo{
S^1\times S^1\times\ldots \times S^1,
}
where each circled action $S^1$ acts on
$\G(W^+_\eufm{c_i}\otimes E_i)$
by the unit modulus, i.e,
\eqn\yya{
q^i \rightarrow e^{i\theta_i} q^i,\qquad i=1,\ldots, N_f.
}
However, if the sequence of the representations
$(R_1,\ldots, R_{N_f})$ contains the same representation
the above global symmetry should be
enhanced accordingly.
This problem can be avoided by adding
the bare mass terms for each
hypermultiplets to maintain the
above symmetry only.
Since this is also the symmetry of the
theory, one can mod out this circled action  to reduce
the path integral   to
the quotient space
\eqn\quot{
\CM(k,\eufm{c_1},\ldots,\eufm{c}_{N_f})/T^{N_f}.
}

The above quotient space has very rich structures.
For example, we consider a particular
global phase symmetry
\eqn\yyb{
q^{N_f } \rightarrow e^{i\theta} q^{N_f}.
}
We have a natural
fixed point of the action for $q^{N_f}=0$. The fixed point
is the moduli space $\CM(k,\eufm{c_1},\ldots,\eufm{c}_{N_f-1})$
which characterizes the theory $T(\eufm{c_1},\ldots,\eufm{c}_{N_f-1} )$.
The other fixed point of the action with $q^{N_f}\neq 0$
is in the (abelian) Coulomb phase.
If there is a non-zero solution $\phi$,
the gauge symmetry breaks down to its maximal torus.
Then, it is always possible to find fixed point of the $S^1$
action by a suitable choice of $q^{N_f}$ in the representation
$R_{N_f}$. By considering the various $S^1$ actions together,
one can find many different fixed points which
include $\CM(k), \CM(k, \eufm{c}_1),\ldots,
\CM(k,\eufm{c}_1,\dots,\eufm{c}_{N_f-1})$.
In the abelian Coulomb phase, we also have a variety of
the fixed points.
It turns out that one can design a version of TQCD
such that the path integral is localized to those
various fixed points.  It is also  possible
to express the topological quantities
defined by the evaluation of the Donaldson
map on the moduli space
$\CM(k,\eufm{c_1},\ldots,\eufm{c}_{i})$
as the sum of contributions of the fixed points in the
Coulomb phase \HPPb.
Thus, we can obtain a dual description of the various theories
$T(\eufm{c_1},\ldots,\eufm{c}_{N_i} )$by examining
the bigger theory.

The above picture suggests that we should reexamine the implication
of the electro-magnetic duality of the $N=2$ supersymmetric QCD.
Naively speaking,  an asymptotically free
$N=2$ supersymmetric QCD,
$SQC\!D(N_f)$,  and its dual theory can be embedded
in another asymptotically free
theory, $SQC\!D(N_f +1)$, coupled with additional matter multiplet.
More precisely, the weakly coupled ultraviolet limit
and the strongly coupled
infrared limit of   $SQC\!D(N_f)$ are realized as certain
two-different fixed points
of the weakly coupled ultraviolet limit of $SQC\!D(N_f +1)$.
Clearly,
$SQC\!D(N_f +1)$ has  milder scaling dependency
than $SQC\!D(N_f)$
as the quantum field theory. To know the strongly coupled
infrared limit of   $SQC\!D(N_f+1)$, one should examine
the weakly coupled ultraviolet limit of $SQC\!D(N_f+2)$
which again has  milder scaling dependency.
Finally, we can end up with the critical theory $SQC\!D(N_{f_{\!c}})$
which does not have any dependency on scaling.
This critical theory, then, should be a self-dual theory.
Thus, the ultimate solutions of the problems amount
to solving all the critical theories.  The existence of
the electro-magnetic dual description of the asymptotically free $N=2$
supersymmetric QCD is originated from the self-duality or
the quantum scaling invariance  of the critical theory.

The self-duality of the critical theory was
suggested by Witten \WittenN\ and also by Seiberg
in the context of the $N=1$ theory \SBb.
 It will be also interesting to see if our picture on the $N=2$
supersymmetric gauge
theories has an application in the $N=1$ theories \SBb
( for a survey and further references see\SBc).
The details will be discussed elsewhere.

\ack{
We would like to thank T.~Mrowka for an useful communication.
SH and JSP would like to thank the organizers of the Jerusalem
Winter School for hospitality.
The work of JP is supported by U.S. Department of Energy
Grant No. DE-FG03-92-ER40701.
JP would like to thank the Isaac Newton Institute for hospitality.
}

\listrefs
\bye